\pdfoutput=1
\documentclass[english]{elsarticle}
\usepackage[T1]{fontenc}
\usepackage[latin9]{inputenc}
\usepackage{units}
\usepackage{graphicx}



\usepackage{babel}
\begin{document}

\title{Suppressing the Numerical Cherenkov Instability in FDTD PIC Codes}

\author{Brendan B. Godfrey}

\address{University of Maryland, College Park, Maryland 20742, USA and Lawrence
Berkeley National Laboratory, Berkeley, California 94720, USA}

\author{Jean-Luc Vay}

\address{Lawrence Berkeley National Laboratory, Berkeley, California 94720,
USA}
\maketitle

\section{Introduction}

The numerical Cherenkov instability \citep{godfrey1974numerical}
is the most serious numerical instability affecting multidimensional
particle-in-cell (PIC) simulations of relativistic particle beams
and streaming plasmas \citep{VayJCP2011,Spitkovsky:ICNSP2011,Xu2013}.
It arises from coupling between numerically distorted electromagnetic
modes and spurious beam modes, the latter due to the mismatch between
the Lagrangian treatment of particles and the Eulerian treatment of
fields \citep{godfrey1975canonical}.

In recent papers we derived and solved electromagnetic dispersion
relations for the numerical Cherenkov instability for both finite
difference time-domain (FDTD) \citep{godfrey2013esirkepov} and pseudo-spectral
time-domain (PSTD) \citep{Godfrey2013PSATD,Godfrey2013PSATD-TPS}
algorithms and successfully compared results with those of the Warp
simulation code \citep{Warp}. From the FDTD analysis we developed
approximate analytical growth rate expressions for the numerical Cherenkov
instability and from them explained the previously observed ``magic
time-steps'' \citep{VayJCP2011,Xu2013} at which instability growth
rates decreased significantly. Our PSTD analysis, focused on Haber's
Pseudo-Spectral Analytical Time-Domain (PSATD) algorithm \citep{HaberICNSP73,Vay2013PSATD},
provided several methods for suppressing the numerical Cherenkov instability.
This was accomplished by a combination of digital filtering at large
wave-numbers and improved numerical balancing of transverse electric
and magnetic fields at smaller wave numbers. Doing either or both
is, of course, mechanically straightforward for PSTD algorithms, because
the currents and fields are known in Fourier space and, therefore,
can be rescaled easily by the desired k-dependent factors.

In this brief paper we demonstrate that the same can be done economically
and with acceptable accuracy for FDTD algorithms without resorting
to Fourier transforms. The stabilization process is described in Sec.
2, numerical solutions presented in Sec. 3, and sample simulation
results given in Sec. 4. Sec. 5 summarizes the paper and suggests
further investigations.

\section{Stabilization procedure}

Ref. \citep{godfrey2013esirkepov} derives the FDTD dispersion relation
for multidimensional PIC codes employing the Esirkepov current-conserving
algorithm \citep{esirkepov2001exact}. In the high energy limit it
can be written as
\begin{equation}
C_{0}+n\sum_{m_{z}}C_{1}\csc\left[\left(\omega-k_{z}^{\prime}\right)\frac{\Delta t}{2}\right]+n\sum_{m_{z}}C_{2}\csc^{2}\left[\left(\omega-k_{z}^{\prime}\right)\frac{\Delta t}{2}\right]=0,\label{eq:drform}
\end{equation}
with coefficients $C_{0},\: C_{1},\: C_{2}$ defined in Eqs. (29)
- (31) of \citep{godfrey2013esirkepov}. Eq. (\ref{eq:drform}) involves
sums over numerical aliases, $k_{z}^{\prime}=k_{z}+m_{z}\,2\pi/\Delta z$
, for wave numbers aligned with the direction, \emph{z}, of beam propagation.
In the limit of vanishingly small time-steps and cell-sizes, Eq. (\ref{eq:drform})
simplifies to $C_{0}=n$, as expected. Thus, all beam resonances in
Eq. (\ref{eq:drform}) are numerical artifacts, even $m_{z}=0$. Their
interaction with the light modes represented by $C_{0}$ gives rise
to the numerical Cherenkov instability. Almost always, the $m_{z}=0,\,-1$
modes dominate. As noted in the Introduction, instabilities are fastest
growing at resonances, which typically occur at large wave numbers
(see Figs. 2 and 6 of \citep{godfrey2013esirkepov}), where they can
be eliminated by digital filtering of the sort described in \citep{VayJCP2011}.

The non-resonant form of the numerical Cherenkov instability, although
slower growing, is more troublesome, because it often occurs at smaller
wave numbers where physical phenomena of interest also occur. We found
when analyzing PSATD algorithms that multiplying the fields as seen
by the particles at small wave-numbers by k-dependent factors differing
from unity by only a few percent was sufficient to cancel the numerical
mismatch in $C_{2}$ between transverse electric and magnetic fields
and, thereby, nearly eliminate the numerical Cherenkov instability
\citep{Godfrey2013PSATD-TPS}. In principle, the same can be done
for FDTD PIC codes. For instance, choosing the multiplier for $E_{x}$
to equal
\begin{equation}
\left.\frac{S^{B_{y}}}{[k_{z}]}/\frac{S^{E_{x}}}{\left[\omega\right]}\right|_{\omega=k_{z}}\label{eq:eb1}
\end{equation}
causes $C_{2}$ for $m_{z}=0$ to vanish at $\omega=k_{z}$. Various
quantities in Eq. (\ref{eq:eb1}) are defined in \citep{godfrey2013esirkepov}.
Inserting those definitions into Eq. (\ref{eq:eb1}) yields
\begin{equation}
\sin\left(k_{z}\frac{\Delta t}{2}\right)\cos\left(k_{z}\frac{\Delta t}{2}\right)\csc^{2}\left(k_{z}\frac{\Delta z}{2}\right)\left(k_{z}\frac{\Delta z}{2}\right)\left(\frac{\Delta z}{\Delta t}\right)\label{eq:Galerkin}
\end{equation}
for the Galerkin interpolation scheme in \citep{godfrey2013esirkepov},
\begin{equation}
\sin\left(k_{z}\frac{\Delta t}{2}\right)\cos\left(k_{z}\frac{\Delta t}{2}\right)\csc\left(k_{z}\frac{\Delta z}{2}\right)\left(\frac{\Delta z}{\Delta t}\right)\label{eq:Uniform}
\end{equation}
for the Uniform interpolation scheme, and
\begin{equation}
\sin\left(k_{z}\frac{\Delta t}{2}\right)\cos\left(k_{z}\frac{\Delta t}{2}\right)\cot\left(k_{z}\frac{\Delta z}{2}\right)\left(\frac{\Delta z}{\Delta t}\right)\label{eq:Momentum}
\end{equation}
for a ``momentum-conserving'' interpolation scheme in which fields
are averaged from the usual Yee staggered mesh \citep{Yee} to a non-staggered
mesh before interpolation to the particles.

\begin{figure}
\centering{}\includegraphics[scale=0.69]{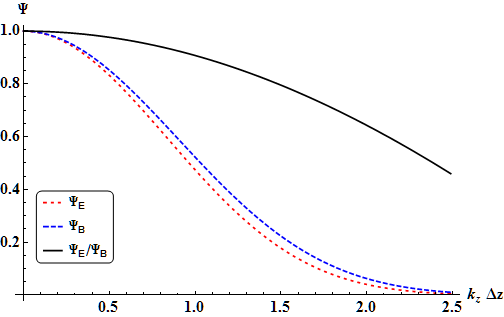}\caption{\label{fig:Field-mul}Field multipliers $\Psi$ for Uniform interpolation
scheme with $\Delta t/\Delta z=0.9$.}
\end{figure}

To apply the multipliers without resorting to Fourier transforms,
they are approximated by ratios of fourth-order polynomials in $\sin^{2}\left(k_{z}\Delta z/2\right)$,\emph{
i.e.}, by rational interpolation functions \citep{Larkin1967}. The
variable $\sin^{2}\left(k_{z}\Delta z/2\right)$ is chosen for its
simple (1, -2, 1) stencil in \emph{z}, and fourth-order is chosen
as a reasonable compromise between accuracy and economy. The nine
interpolation points from which the polynomial coefficients are determined
are spaced uniformly between 0 and 1 in $\sin^{2}\left(k_{z}\Delta z/2\right)$
. Better choices for the interpolation points may exist. As in the
earlier papers, Mathematica \citep{Mathematica9} is used to solve
the dispersion relation; its RationalInterpolation function conveniently
provides the desired polynomial coefficients.%
\footnote{Software to calculate these coefficients is available in Computable
Document Format \citep{WolframCDF} at http://hifweb.lbl.gov/public/BLAST/Godfrey/.%
} Fig. \ref{fig:Field-mul} displays $\Psi_{E}$, the numerator polynomial,
$\Psi_{B}$, the denominator polynomial, and the ratio of the two,
the rational interpolation function for, in this case, Eq. \ref{eq:Uniform}
with $\Delta t/\Delta z=0.9$ and $\beta_{z}=1/8$ (\emph{i.e.}, with
the Cole-Karkkainen field-solver \citep{ColeIEEE1997,ColeIEEE2002,KarkICAP06}).
The rational interpolation function is accurate to $10^{-6}$ except
for the largest values of $k_{z}\Delta z$ shown in the figure, where
the accuracy still is better than $10^{-5}$.

The numerator, $\Psi_{E}$, of the rational interpolation function
can be applied to $E_{x}$ before interpolation to the particles by
repeated applications of the (1, -2, 1) stencil, and the denominator,$\Psi_{B}$,
by repeated application of its inverse, in other words, by tri-diagonal
matrix inversion. However, because some digital filtering is needed
anyway, it is simplier to apply only the numerator polynomial to $E_{x}$,
and the denominator polynomial to $B_{y}$ and $E_{z}$, in effect
using the denominator as a digital filter. (In three dimensiions $\Psi_{E}$
is applied to both transverse \emph{E}-fields, and $\Psi_{B}$ to
all other fields.) This approach is implemented in WARP.

\section{Numerical solutions}

This section presents solutions of the complete FDTD-Esirkepov linear
dispersion relation, Eq. (11) of \citep{godfrey2013esirkepov} with
interpolation function $S^{E_{x}}$ multiplied by $\Psi_{E}$, and
$S^{B_{y}}$ and $S^{E_{z}}$ multiplied by $\Psi_{B}$. Digital filtering
is as in Eq. (37) of \citep{godfrey2013esirkepov} but with the exponent
``16'' replaced by ``4'', 
\begin{equation}
\cos^{4}\left(k_{z}\frac{\Delta z}{2}\right)\left(5-4\cos^{2}\left(k_{z}\frac{\Delta z}{2}\right)\right)^{2}\cos^{4}\left(k_{x}\frac{\Delta x}{2}\right)\left(5-4\cos^{2}\left(k_{x}\frac{\Delta x}{2}\right)\right)^{2},\label{eq:filter}
\end{equation}
equivalent to two passes each of a bilinear filter and a compensation
filter \citep{VayJCP2011}. Thus, much less digital filtering is applied.
Cubic interpolation is employed on a two-dimension, rectangular, periodic,
128x128 mesh with cell size 0.3868. Linear interpolation also was
tried but did not produce uniformly good stability results.

\begin{figure}
\begin{centering}
\includegraphics[scale=0.95]{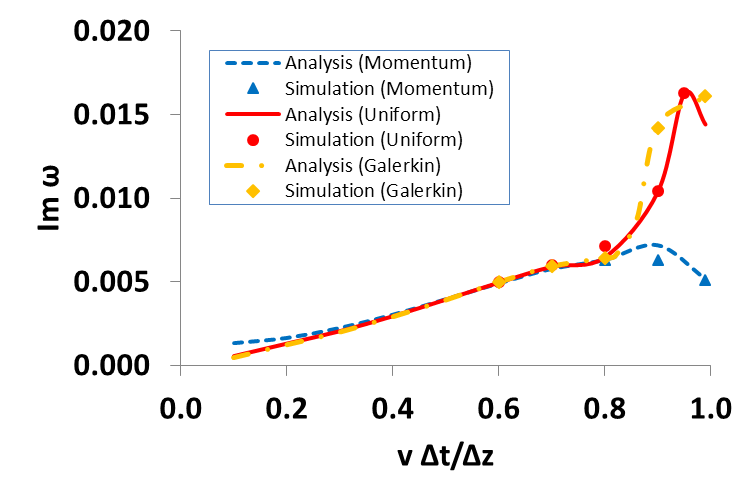}
\par\end{centering}

\caption{\label{fig:g130}Maximum numerical instability growth rates observed
in WARP and calculated from the numerical dispersion relation for
$\gamma=130.$}
\end{figure}

\begin{figure}
\centering{}\includegraphics[scale=0.95]{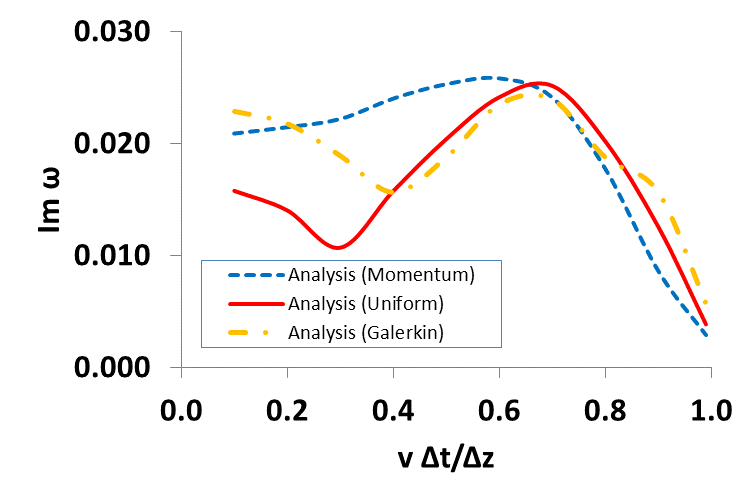}\caption{\label{fig:g3}Maximum numerical instability growth rates calculated
from the numerical dispersion relation for $\gamma=3.$}
\end{figure}

Maximum numerical instability growth rates for a beam of energy $\gamma=130$
are shown as a function of $v\,\Delta t/\Delta z$ in Fig. \ref{fig:g130}.
(Note that the determination of $\Psi_{E}$ and $\Psi_{B}$ breaks
down at $0.756<v\,\Delta t/\Delta z<0.764$, and this narrow region
has been excised from the Figs. \ref{fig:g130} and \ref{fig:g3}.)
Also shown are growth rates obtained from WARP simulations; agreement
is good. These growth rates are much smaller than those obtained without
the rational interpolation factors but with otherwise identical parameters,
given by the curves labeled ``Galerkin-CK'' and ``Uniform-CK''
in Fig. 16 of \citep{Godfrey2013PSATD}. In fact, the growth rates
in Fig. \ref{fig:g130} are comparable to those labeled ``PSATD (c)''
in Fig. 16 of \citep{Godfrey2013PSATD}, which is not surprising:
Both are based on zeroing C2 at $\omega=k_{z}$. Although not shown
here, the dispersion relation also has been solved for variants of
the Fig. \ref{fig:g130} parameters with (1) no digital filtering
(apart from $\Psi_{B}$), which roughly doubles growth rates at large
$v\,\Delta t/\Delta z$; (2) $\Psi_{E}$/$\Psi_{B}$ applied to $E_{x}$
only, which roughly triples growth rates at large $v\,\Delta t/\Delta z$;
and (c) not employing the compensation filter, which moderately reduces
growth rates, especially at large $v\,\Delta t/\Delta z$. Of course,
physical results at small wave numbers also may be filtered modestly
in this last case \citep{VayJCP2011}.

Because the scaling procedure described in this paper is tuned for
infinite $\gamma$, it works less well at only modestly relativistic
beams, as illustrated in Fig. \ref{fig:g3} for $\gamma=3$. Based
on this and other computations, peak growth rates appear to scale
here very roughly as $\gamma^{-\nicefrac{1}{2}}$. Tuning the procedure
for moderate $\gamma$ is more difficult, because the dispersion relation
is more complicated. Nonetheless, the numerical instability growth
rates depicted in Fig. \ref{fig:g3} probably are acceptable for most
purposes. At still smaller beam energies, the well known $m_{z}=-1$
quasi-one-dimensional, electrostatic numerical instability \citep{Langdon1970,Okuda1970}
dominates. It can be suppressed by using any field interpolation algorithm
that offsets $E_{z}$ by $\triangle z/2$ relative to the charge density
$\varrho$ (or $\mathbf{W}$ in the Esirkepov current algorithm) and
interpolates it with a spline one order lower in z relative to $\varrho$
or to $\mathbf{W}$ \citep{lewis1972variational,langdon1973energy},
such as the Galerkin algorithm.

\section{Application to laser plasma acceleration simulations}

\begin{figure}
\centering{}\includegraphics[scale=0.47]{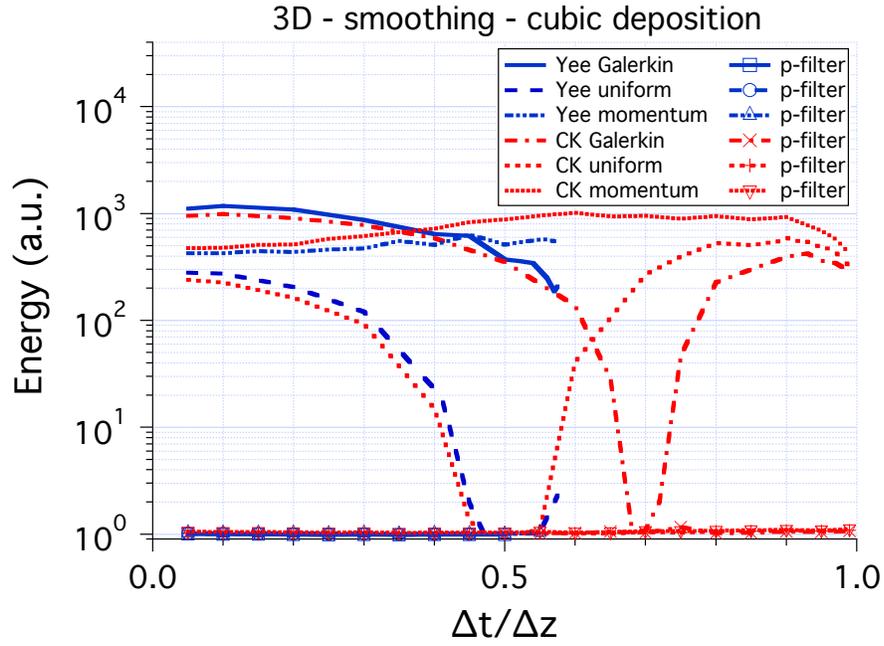}\caption{\label{fig:WARPscan3D}Field energy relative to stable reference level
vs $\Delta t/\Delta z$ from three dimensional WARP LPA simulations
at $\gamma$ = 13, using Galerkin, uniform, and momentum-conserving
field interpolation with either standard (``Yee'') or Cole-Karkainnen
(``CK'') field solvers. The relative energy is essentially unity
for all instances in which the rational interpolation function (``p-filter'')
is applied.}
\end{figure}

In a laser plasma accelerator (LPA), a laser pulse is propagated through
a plasma, creating a wake of very strong electric fields of alternating
polarity \citep{TajimaPRL79}. An electron beam injected with the
appropriate phase thus can be accelerated to high energy in a distance
much shorter than those for conventional acceleration techniques \citep{LeemansNature06}.
As a verification that the theory developed in this paper applies
to the modeling of LPAs, series of three dimensional simulations of
a 100 MeV class LPA stage were performed, focusing on the plasma wake
formation. The velocity of the wake in the plasma corresponds to $\gamma\simeq13.2$,
and the simulations were performed in a boosted frame of $\gamma_{f}=13.$
The LPA simulations with parameters leading to Figs. 15 and 16 in
\citep{godfrey2013esirkepov} were repeated using the procedure discussed
in this paper, providing results such as those in Fig. \ref{fig:WARPscan3D}.
(Note that both these and the simulation sweeps in \citep{godfrey2013esirkepov}
used the digital filter with one pass each of a bilinear filter and
a compensation filter, equivalent to the square root of the expression
in Eq. \ref{eq:filter}, rather than the expression in Eq. (37) of
\citep{godfrey2013esirkepov}.) Energy conservation was excellent
when the rational interpolation multipliers were used.

\begin{figure}
\centering{}\includegraphics[scale=0.5]{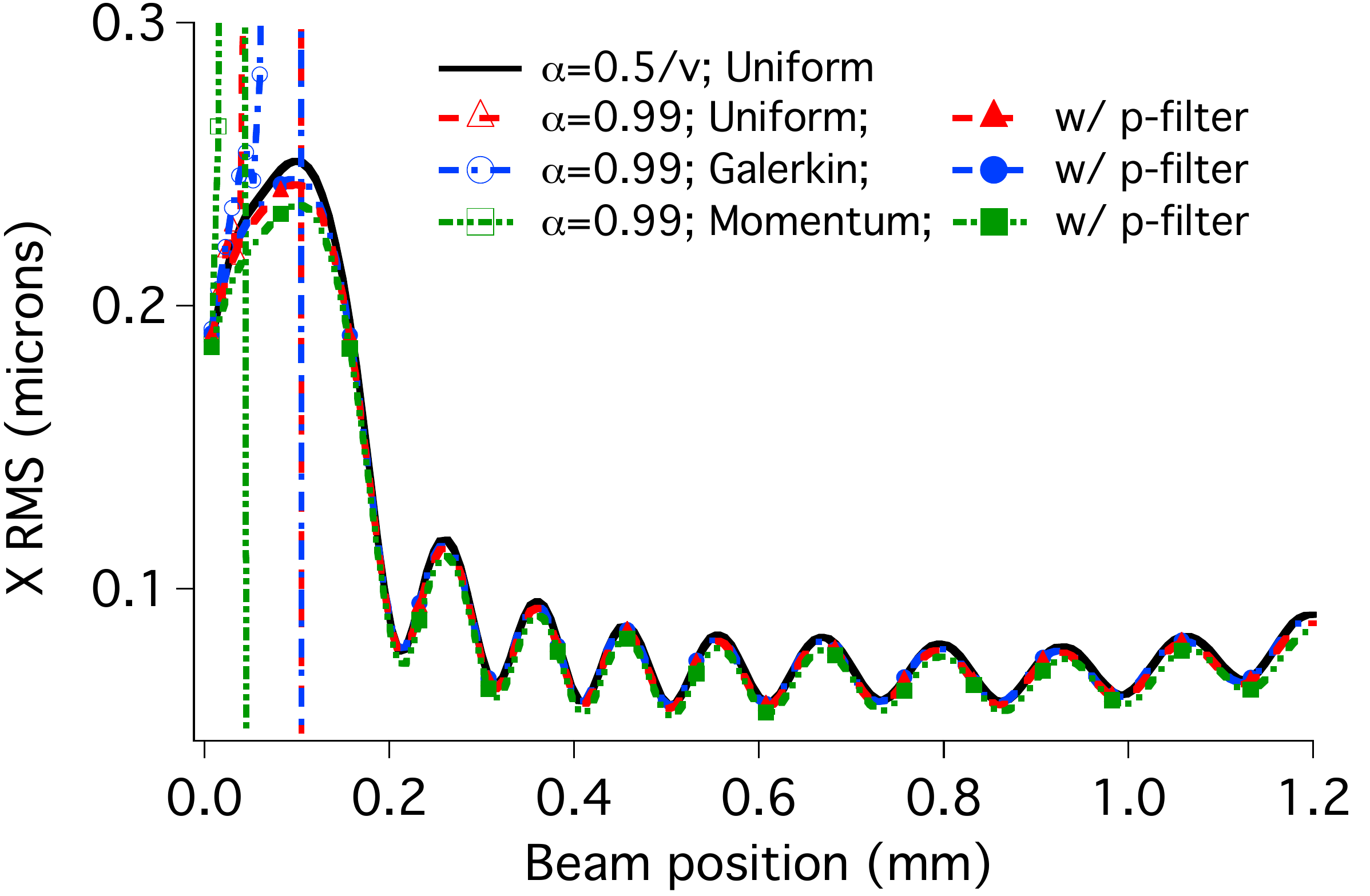}\caption{\label{fig:LPA_Xrms}Accelerated electron beam RMS radius from $\gamma=13$
LPA simulations. Six runs used $\alpha\equiv\Delta t/\Delta z=0.99$;
those employing the rational interplation multipliers (``p-filter'')
behaved well, while those that did not produced meaningless results.
An almost instability-free uniform interpolation simulation at the
$v\,\Delta t/\Delta z$=0.5 ``magic time step'' is provided for
comparison.}
\end{figure}

Additionally, three dimensional $\gamma=13$ LPA simulations of the
sort described in \citep{VayPoP2011} were performed to validate the
stabilization procedure. Fig. \ref{fig:LPA_Xrms} records the accelerated
electron beam RMS radius. Six simulations used $\Delta t/\Delta z=0.99$;
those employing the rational interpolation multipliers behaved as
desired, while those that did not produced meaningless results due
to the numerical Cherenkov instability. An almost instability-free
uniform interpolation simulation at the $v\,\Delta t/\Delta z$=0.5
``magic time step'' is provided for comparison. All four stable
simulations provided essentially identical results, including the
evolution of beam energy (not shown), emittance (not shown), and radius.

\section{Conclusion}

This paper presents a straightforward approach for greatly reducing
numerical Cherenkov instability growth rates in FDTD-Esirkepov PIC
simulations of relativistic beams and streaming plasmas. Moreover,
sample simulations indicate that this approach is economical, requires
minimal additional digital filtering, and apparently has no adverse
effect on physical results at wavelengths long compare to the simulation
axial cell size. Although derived for highly relativistic flows, it
works reasonably well down to $\gamma$ of order 3, below which the
numerical Cherenkov instability ceases to be the dominant numerical
effect.

While this approach seems quite promising, possibly even better procedures
may be possible, including some adaptable from approaches already
demonstrated for PSATD PIC algorithms \citep{Godfrey2013PSATD,Godfrey2013PSATD-TPS}.
We hope to explore such alternatives in the near future. We also anticipate
generalizing these procedures to FDTD simulations not employing the
Esirkepov algorithm, and to a wider range of Maxwell solvers.

\section*{Acknowledgments}

We wish to thank Irving Haber for suggesting this collaboration and
for helpful recommendations. We also are indebted to David Grote for
assistance in using the code WARP. This work was supported in part
by the Director, Office of Science, Office of High Energy Physics,
U.S. Dept. of Energy under Contract No. DE-AC02-05CH11231 and the
US-DOE SciDAC ComPASS collaboration, and used resources of the National
Energy Research Scientific Computing Center.

This document was prepared as an account of work sponsored in part
by the United States Government. While this document is believed to
contain correct information, neither the United States Government
nor any agency thereof, nor The Regents of the University of California,
nor any of their employees, nor the authors makes any warranty, express
or implied, or assumes any legal responsibility for the accuracy,
completeness, or usefulness of any information, apparatus, product,
or process disclosed, or represents that its use would not infringe
privately owned rights. Reference herein to any specific commercial
product, process, or service by its trade name, trademark, manufacturer,
or otherwise, does not necessarily constitute or imply its endorsement,
recommendation, or favoring by the United States Government or any
agency thereof, or The Regents of the University of California. The
views and opinions of authors expressed herein do not necessarily
state or reflect those of the United States Government or any agency
thereof or The Regents of the University of California.

\section*{References}

\bibliographystyle{elsarticle-num}
\bibliography{C:/Users/Brendan/Documents/LyX/Biblio_JCP_Godfrey,C:/Users/Brendan/Documents/LyX/Godfrey}

\clearpage{}

\clearpage{}

\clearpage{}

\clearpage{}

\end{document}